\documentclass[
superscriptaddress,
 amsmath,amssymb,
 aps,
 prl,
 lengthcheck,%
]{revtex4}

\usepackage{graphicx}
\usepackage{dcolumn}
\usepackage{bm}
\usepackage{hyperref}
\usepackage{feynmp}
\usepackage{amsmath}

\begin{document}


\title{Real-Time Ginzburg-Landau Theory for Bosons in Optical Lattices}

\author{T. D. Gra\ss}
\author{F. E. A. dos Santos}
\affiliation{Institut f\"ur Theoretische Physik, Freie Universit\"at Berlin, Arnimallee 14, 14195 Berlin, Germany}
\author{A. Pelster}
\affiliation{Institut f\"ur Physik und Astronomie, Universit\"at Potsdam, Karl-Liebknecht-Stra\ss e 24/25, 14476 Golm, Germany}
\affiliation{Fachbereich Physik, Universit\"at Duisburg-Essen, Lotharstra\ss e 1, 47048 Duisburg, Germany}

\date{\today}

\begin{abstract}
Within the Schwinger-Keldysh formalism we derive a Ginzburg-Landau theory for the Bose-Hubbard model which describes the real-time dynamics of the complex order parameter field. Analyzing the 
excitations in the vicinity of the quantum phase transitions it turns out that particle/hole dispersions in the Mott phase map continuously onto corresponding amplitude/phase excitations in the 
superfluid phase. Furthermore, in the superfluid phase we find a sound mode, which is in accordance with recent Bragg spectroscopy measurements in the
Bogoliubov regime, as well as an additional gapped mode, 
which seems to have been detected via lattice modulation.
\end{abstract}


\maketitle

\newcommand{\op}[1]{\ensuremath{\Hat{\mathrm{#1}}}}

Within the last decade ultracold atoms in optical lattices \cite{jaksch,bloch1} have become a standard tool for studying quantum-statistical many-body effects. Due to their high tunability, these 
systems represent an almost 
perfect test ground for a large variety of solid-state models. In particular, the experimental observation of the seminal quantum phase transition from the Mott insulating (MI) 
to the superfluid (SF) phase, exhibited by a single-band Bose-Hubbard (BH) system of spinless or spin-polarized bosons, has recently attracted a lot of attention \cite{bloch-review}. Although the 
occurrence of this quantum phase transition is evident from the momentum distributions of destructive time-of-flight measurements, its precise location cannot be determined from them. Recently, 
however, more detailed information about the collective excitations of this system could also be achieved in a non-destructive way by exciting the system via lattice modulation \cite{esslinger} 
or by Bragg spectroscopy \cite{sengstock}. Deep in the SF phase, the observed gapless excitation spectrum can be well understood within a Bogoliubov theory \cite{stoof}. Approaching the phase boundary, 
both a slave-boson method \cite{blatter1}  and a random-phase approximation \cite{menotti} have even predicted an additional SF gapped mode. When the MI phase is reached, both SF modes turn continuously 
into particle and hole excitations which are also found by mean-field theory \cite{stoof}. By 
being both gapped, they characterize the insulating phase in a unique way. Until now, however, there exists no 
theory which describes the collective excitations in these physically different regimes in a unified way \cite{stoofbook}.\\
To this end we derive in this Letter a real-time Ginzburg-Landau theory within the Schwinger-Keldysh formalism. Following Refs. \cite{ednilson,barry} our approach is technically based on resumming a 
perturbative hopping expansion for the effective action. This is physically justified as a dimensional rescaling of the hopping parameter $J$, which reads $J\rightarrow J/D$  for dimensions $D>1$
 \cite{vollhardt} in the presence of a condensate, turns out to suppress all $n$th order hopping loops by a factor $1/D^{n-1}$. Resumming the 1-particle irreducible contributions up to the $n$th 
hopping order yields an effective $(1/D)$-expansion up to the $(n-1)$th order. 
By restricting ourselves in this Letter to the lowest order $n=1$, we will get an approximation which is exact for infinite 
dimensions or infinite-range hopping \cite{fisher} and turns out to describe well the quantum phase transition for $D=2,3$. Although our approach is originally designed for describing the vicinity of 
the phase transition, it turns out to reproduce also the Bogoliubov spectrum deep in the SF phase. This is an evidence that
our Ginzburg-Landau theory can be applied for quantitative predictions within a broad range of 
system parameters.\\
For our perturbative calculation we use the Dirac picture, where the dynamics of the operators is given by an unperturbed Hamiltonian $\op{H}_0$. To this end we introduce the splitting of the 
BH-Hamiltonian $ \op{H}_{\mathrm{BH}} = \op{H}_0 + \op{H}_{\mathrm{kin}}$, where 
$\op{H}_0 = \sum_{i} \left[ \frac{U}{2} \op{a}_i^{\dagger} \op{a}_i \left( \op{a}_i^{\dagger} \op{a}_i -1 \right) - \mu \op{a}_i^{\dagger} \op{a}_i \right]$ is the local interaction and 
$\op{H}_{\mathrm{kin}}= - \sum_{i,j} J_{ij} \op{a}_i^{\dagger} \op{a}_j$ the hopping term. Here $\op{a}_i$ ($\op{a}^\dagger_i$) denote the bosonic annihilation (creation) operators at lattice site 
$i$, $\mu$ the chemical potential, $U$ the on-site interaction parameter, and $J_{ij}$ the hopping matrix element being equal to $J>0$ for nearest neighbors only. In order to break the
underlying $U(1)$ symmetry of the BH-Hamiltonian, we add an artificial 
source term $\op{H}_{\mathrm{S}}(t) = \sum_i \left[ j_i(t) \op{a}_i^\dagger + \mathrm{c.c.} \right]$ to the Hamiltonian with external currents $j_i(t)$. Considering the source term as part of the 
perturbation $\op{H}_1(t) = \op{H}_{\mathrm{kin}} + \op{H}_{\mathrm{S}}(t)$, we arrive at $\op{H}(t)=\op{H}_0 + \op{H}_1(t)$.\\
Within the Schwinger-Keldysh formalism the Dirac picture involves a time evolution along a closed real-time contour which starts and ends at some initial time $t_0$ and must extend to any large time 
which can be chosen to be $+ \infty$. To define the position of the operators on the contour, we must provide them with an additional path index $\mathrm{P}=\pm$ \cite{kamenev2}. The resulting 
time-evolution operator along the closed contour then reads
\begin{eqnarray}
 \op{S}^\dagger \op{S} \equiv \op{T}_\mathrm{c} \exp \left[ \sum_{\pm} \mp \frac{i}{\hbar} \int_{t_0}^{\infty} \mathrm{d}t' \ \op{H}_{1\pm}(t') \right],
\end{eqnarray}
where the contour-ordering operator $\op{T}_{\mathrm{c}}$ turns the operators of the forward path time-ordered, followed by the anti-time-ordered operators of the backward branch. 
If we distinguish accordingly the currents $j$ on the forward path of the contour from the backward ones, the resulting generating functional 
$ {\cal Z}[j,j^*] = \left\langle \op{T}_{\mathrm{c}} \{ \op{S}^\dagger \op{S} \} \right\rangle_0$ defines the corresponding Green's functions.
Denoting the set of variables $\{i,t,\mathrm{P}\}$ by Greek indices, the contour-ordered Green's functions read:
\begin{eqnarray}
\label{HD} 
&& G_{\alpha_1,\cdots,\alpha_n;\alpha_{n+1},\cdots,\alpha_{n+m}} \equiv i^{n+m-1}
 \\ && \times \nonumber
\Big\langle \op{T}_\mathrm{c} \Big\{ \op{S}^\dagger \op{S}
\op{a}_{\alpha_1}\cdots \op{a}_{\alpha_n} \op{a}^\dagger_{\alpha_{n+1}} \cdots \op{a}^\dagger_{\alpha_{n+m}} \Big\} \Big\rangle_0.
\end{eqnarray}
The angle brackets $\langle \cdot \rangle_0$ denote a thermal average with respect to the unperturbed Hamiltonian $\op{H}_0$. In order to account for perturbative contributions to the thermal average, 
one should actually choose a time-contour which also consists of an imaginary part from $t_0$ to $t_0 -i \hbar \beta$. However, it is widely believed in the literature that, if we push 
$t_0\rightarrow-\infty$, this imaginary part can be neglected for initially uncorrelated systems  \cite{chou,rammer-buch}. We will follow this tradition and discuss its limitations at the end 
of this Letter.\\
The coefficients of an expansion of the generating functional $\cal Z$ with respect to both the currents and the hopping parameter are given in terms of \textit{unperturbed} Green's functions. A 
simple diagrammatic construction rule for the expansion is found, when we consider the functional ${\cal F}[j,j^*] = -i \ln {\cal Z}[j,j^*]$. According to the linked-cluster the expansion coefficients 
for this functional are the cumulants or connected Green's functions \cite{singh, metzner}, which are related to the Green's functions (\ref{HD}) by decomposition formulas. Defining 
$j_\alpha \equiv \mathrm{P}_\alpha j_{i_\alpha\mathrm{P}_\alpha}(t_\alpha)/\hbar$,  where the sign $\mathrm{P}_{\alpha}$ takes into account the direction of the time evolution, 
and $J_{\alpha\beta} \equiv J_{i_\alpha i_\beta}/\hbar$, we get in fourth order in $j$ and first order in $J$:
\begin{align}
\label{F}
& {\cal F}= j_{\alpha} \left[ C_{\alpha\beta} + \mathrm{P}_\kappa J_{\alpha\beta} C_{\alpha\kappa}C_{\kappa\beta} \right] j^*_{\beta} + j_\alpha j_\beta 
\\ & \nonumber 
\times
\left[\frac{1}{4} C_{\alpha\beta\gamma\delta} +\mathrm{P}_{\kappa} \frac{1}{2} J_{\beta\gamma} \left(
C_{\alpha\kappa\gamma\delta} C_{\beta\kappa} + \mathrm{h.c.} \right) \right] j^*_\gamma j^*_\delta.
\end{align}
Doubly occuring indices must be summed (integrated). The above notation obscures the locality of the unperturbed cumulants $C$ in its spatial variables. The general structure of Eq. (\ref{F}) 
does not change in Fourier space with the variables $\{\boldsymbol{k}, \omega, \mathrm{P}\}$, if we define $J_{\alpha\beta}\equiv \delta_{\boldsymbol{k}_\alpha,\boldsymbol{k}_\beta}J/\hbar$. Note that the 
frequency is conserved at each cumulant, when transformed in Fourier space.\\
The relation $\hbar \delta {\cal F}/\delta j^*_\alpha = \langle \op{a}_\alpha \rangle$, which defines the order parameter field $\Psi_\alpha$, motivates to perform a Legendre transformation:
\begin{eqnarray}
\label{legendre}
&& \Gamma[\Psi,\Psi^*] \equiv {\cal F}\left[j,j^*\right] - 
\frac{1}{\hbar} \left( j_\alpha \Psi^*_\alpha + \mathrm{c.c.} \right).
\end{eqnarray}
In order to calculate the effective action $\Gamma$ as a power series in $J, \Psi$ and $\Psi^*$, we must invert $\Psi[j,j^*]$ iteratively in both the order parameter fields and the hopping. For 
describing the symmetry-broken behavior, the Ginzburg-Landau theory demands a fourth-order term in $\Psi$. Already a first-order expansion in the hopping yields mean-field results, since all 
diagrams without lattice loops turn out to be resummed due to the Legendre transformation. This can be seen by noticing that the second-order term in Eq. (\ref{F}) is local in Fourier space and 
the only degree of freedom, which remains to be summed, are the path indices. We can get rid of these sums by defining vector currents $\vec{j} \equiv (j_+,-j_-)$ and interpreting the 2-point 
cumulants as 2x2 matrices $C \equiv C_{\mathrm{P}_1\mathrm{P}_2}$. This then allows to write the second-order term as $\vec{j}(C+J C\sigma^3C)\vec{j}^*$ with the Pauli matrix $\sigma^3$ taking into 
account the sign $\mathrm{P}$. The subsequent Legendre transformation then leads to an iterative inversion of this coefficient yielding the inverse of the geometric series $C \sum_n (J \sigma^3 C)^n$, 
which contains all hopping ``chains''.\\
From the resulting Ginzburg-Landau functional
\begin{eqnarray}
\label{Gamma}
\Gamma = && \Psi_\alpha \left[ C^{-1}_{\alpha\beta} - J_{\alpha\beta} \delta_{\mathrm{P}_\alpha,\mathrm{P}_\beta} \mathrm{P}_\alpha \right]\Psi^*_\beta
\\ && \nonumber 
-\frac{1}{4} C_{\alpha\beta\gamma\delta}
C^{-1}_{\alpha\alpha'}\Psi_{\alpha'}
C^{-1}_{\beta\beta'}\Psi_{\beta'}
C^{-1}_{\gamma'\gamma}\Psi^*_{\gamma'}
C^{-1}_{\delta\delta'}\Psi^*_{\delta'},
\end{eqnarray}
we finally obtain equations of motion via extremization, since the currents in the original physical system vanish:
\begin{eqnarray}
\label{eqm}
j^*_\alpha=\frac{\delta \Gamma}{\delta \Psi_\alpha} \stackrel{!}{=}0.
\end{eqnarray}
Together with the complex conjugate of Eq. (\ref{eqm}), there are in total four equations of motion, which simplify when the path-ordered quantities in $\Gamma$ are rotated into the so-called Keldysh 
basis. Here one considers the linear combinations $X_{\Sigma}(t) = (X_+(t) + X_-(t))/\sqrt{2}$ and $X_{\Delta}(t) = (X_+(t) - X_-(t))/\sqrt{2}$.
Then the equation of motion, where the derivative is taken with respect to $\Psi_\Sigma$, can be solved by assuming $\Psi_\Delta=0$, which implies $\Psi_+=\Psi_-$. It turns out that the other equation 
of motion determining $\Psi_\Sigma$ depends only on the retarded and advanced cumulants which are defined as the thermal average of multiple commutators times Heaviside step functions (cf. \cite{chou}).\\
\begin{figure*}
\includegraphics[width= 0.95 \textwidth]{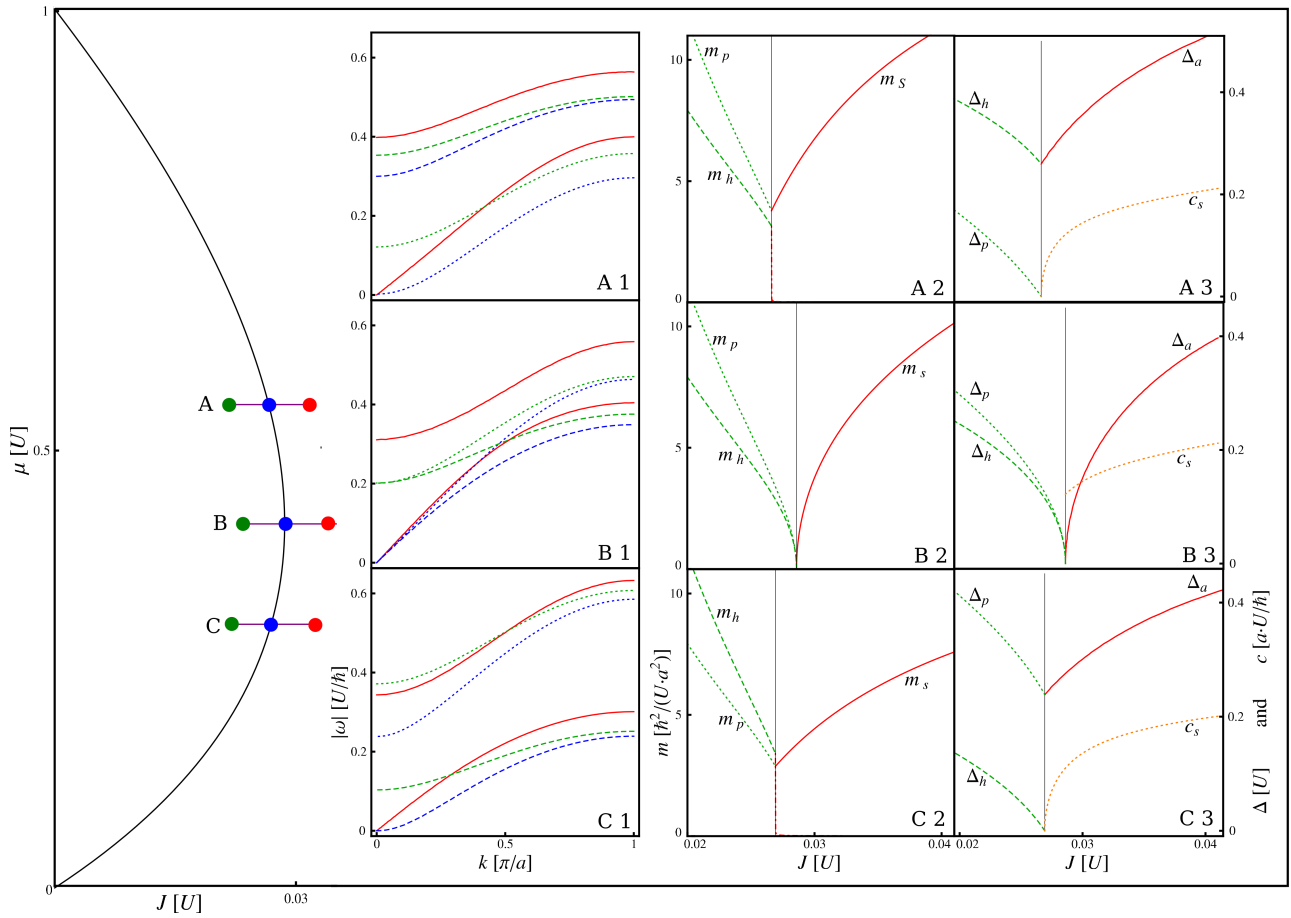}
\caption{\label{spectra} Excitations in $\boldsymbol{k}=(1,0,0)$ direction are plotted in A1 -- C1 for different values $\mu/U$ and $J/U$, which are marked in the phase diagram (left). In the MI 
phase (green) and on the phase boundary (blue), the two $T=0$ modes can be interpreted as particle (dotted) and hole (dashed) excitations. At the tip of the Mott lobe (B), both modes become gapless, 
whereas for larger (smaller) chemical potentials $\mu$ only the gap of the particle (hole) mode vanishes. In the SF phase (red), the gapless mode turns into a sound mode, but also a gapped mode is 
present everywhere in the SF phase. The smooth transition from the MI excitation to the SF excitation is further analyzed in A 2 -- C 2 and A 3 -- C 3, where the effective mass $m$ and the gap $\Delta$ 
of each mode are plotted as a function of $J/U$. The sound velocity $c$ of the massless SF excitation, plotted in A3 -- C3, vanishes at the phase boundary except at the tip, indicating the existence 
of a different universality class in this configuration.}
\end{figure*}
The non-trivial solutions of Eq. (\ref{eqm}) and its complex conjugate can be divided into two classes which we work out at $T=0$:\\
1. Static solutions follow from the ansatz $\Psi_{\Sigma \boldsymbol{k}}(\omega) = \delta_{\boldsymbol{k},\boldsymbol{0}} \delta(\omega) \Psi_{\mathrm{eq}}$. Eq. (\ref{eqm}) then reduces to an algebraic 
equation which determines $\Psi_{\mathrm{eq}}$. By inspecting the region in the parameter space, where 
$\Psi_{\mathrm{eq}}$ starts to become non-vanishing, we find the phase boundary depicted for $T=0$ and $0 \leq \mu/U \leq 1$ on the left side of Fig. \ref{spectra}. In the 
considered first hopping order, this quantum phase diagram is identical to mean-field results \cite{stoof} with a deviation from recent high-precision Monte Carlo data \cite{montecarlo} of about 25\%. 
Within a Landau expansion the second-order hopping contribution has recently been calculated analytically in Ref. \cite{ednilson} decreasing the error down to less than 2\%. A numeric evaluation of 
higher hopping orders has even been shown to converge to a quantum phase diagram which is indistinguishable from the Monte-Carlo result \cite{teichmann2}.\\
2. For obtaining dynamic solutions we perform the ansatz $\Psi_{\Sigma \boldsymbol{k}}(\omega) = \delta_{\boldsymbol{k},\boldsymbol{0}} \delta(\omega) \Psi_{\mathrm{eq}} + \delta \Psi_i(\omega)$ and consider 
only terms which are linear in $\delta \Psi$. With this we obtain algebraic equations which determine the excitation spectra. As the corresponding analytic expressions are rather cumbersome, we restrict 
ourselves to discuss them graphically in Fig.\ref{spectra}.\\
In the MI phase, we find the gapped particle/hole excitations from mean-field theory \cite{stoof}. At the phase boundary, we have to distinguish the tip from the rest of the lobe in accordance with the 
critical theory of the BH model \cite{fisher,sachdev}. While at the tip of the $m$th lobe, i.e. at $\mu_m = \sqrt{m(m+1)}-1$, both excitations become gapless and linear for small $|\boldsymbol{k}|$, 
off the tip at $\mu>\mu_m$ ($\mu<\mu_m$) only the particle (hole) mode becomes gapless and remains with a finite effective mass. In both cases the vanishing of the gap can be described by the exponential 
law $\Delta \sim [J - J_{\rm PB}(\mu)]^{z \nu}$ 
with the mean-field critical exponents $\nu = 1/2$ as well as
$z=1/2$ at the tip and $z=1$ off the tip. Turning into the SF phase, the gapless mode rapidly loses its mass and has to be identified with the Goldstone mode which 
arises due to the broken $U(1)$ symmetry. Indeed, within our Ginzburg-Landau theory it turns out for $\boldsymbol{k}\rightarrow\boldsymbol{0}$ and $\omega\rightarrow0$, that the excitation 
$\delta\Psi$ stems from variations of the phase. Within a slave-boson approach it has even been shown in Ref. \cite{blatter1} that also for general wave vectors $\boldsymbol{k}$ phase variations 
dominate this excitation. This leads to density variations which make this mode sensitive to Bragg spectroscopy. Recently, the whole sound mode has been measured via Bragg spectroscopy far away from the 
phase boundary and could be well described via a Bogoliubov fit \cite{sengstock}. Surprisingly, also this regime deep in the SF phase turns out to be accessible with our Ginzburg-Landau theory. 
Expanding the Green's functions in $U$, in the lowest non-trivial order they do not depend on temperature and lead to the equation of motion
\begin{align}
\label{GPEQ}
 i \hbar \frac{\partial \Psi_i}{\partial t} = -\sum_j J_{ij} \Psi_j -\mu \Psi_i - U \Psi_i |\Psi_i|^2,
\end{align}
which is the lattice version of the Gross-Piteevskii (GP) equation \cite{polkovnikov}. From this follows the Bogoliubov sound mode of a fully condensed system \cite{stoof}. \\
Additionally to that sound mode, however, also a gapped mode survives the quantum phase transition or arises again if we depart from the tricritical point. It can be smoothly mapped onto one of the 
respective MI modes, which is shown in the plot of both the effective masses and the gaps on the right side of Fig.\ref{spectra} The existence of such a SF gapped mode is in accordance with results 
obtained previously in Refs. \cite{blatter1,menotti}, however this finding has not yet been experimentally confirmed. The gapped mode seen via Bragg spectroscopy in Ref. \cite{sengstock} can be well 
explained within a Bogoliubov-de Gennes ansatz considering second-band excitations due to finite temperature. In Ref. \cite{blatter1} it is argued that the physical picture behind the gapped $T=0$ 
mode is an amplitude excitation which corresponds to an exchange between condensed and non-condensed particles at constant overall density. This reasoning is compatible with a numerical investigation 
of our full equations of motion  (\ref{eqm}) where the gapped mode converges for small $U$ to the constant dispersion $\omega(\boldsymbol{k})=2\mu$. Furthermore, due to the absence of any density 
variation, this mode should be insensitive to Bragg spectroscopy. Although Eq. (\ref{eqm}) does not allow \textit{pure} amplitude excitations, we can back this interpretation of predominant 
amplitude excitations by observing that the zero-momentum energy transfer at the phase boundary corresponds to the creation of a particle/hole pair. Via a modulation of the lattice potential, a finite 
energy absorption at zero-momentum transfer has already been observed in Ref. \cite{esslinger}. Although this might be seen as the first experimental signal for such a SF gapped mode, an unambiguous 
quantitative identification is still lacking.\\
Finally we compare our results with the ones obtained within a similar Ginzburg-Landau theory in imaginary time \cite{barry}. Whereas at $T=0$ both formalisms yield identical results, mismatches occur 
for finite temperature. This is surprising, since both formalisms are considered to be equivalent in equilibrium \cite{chou,rammer-buch}. We conclude that the Keldysh formalism working with a purely 
real time-evolution contour is not able to produce the correct equilibrium configuration of the full system for finite temperature. In a non-equilibrium system it might be justified to define the 
temperature only in the unperturbed initial state. For a system, however, that is supposed to relax into a new equilibrium state, a temperature change is expected, if the Hamiltonian describing the 
new equilibrium does not coincide the old one. From this reasoning follows that the imaginary part of the time-evolution contour must not be neglected \cite{evansPRD47}. Furthermore, the agreement of 
both formalisms at $T=0$ can be understood as a consequence of the Gell-Mann-Low theorem \cite{gell-mann} stating that the systems remains in the ground-state, if a perturbation is adiabatically 
switched on. Since our ansatz has pushed this switching into the infinite past, no additional assumptions about its adiabatic properties had to be made \cite{rammer-buch}.\\
The mismatch between our real-time calculation and the previous imaginary-time formalism suggests further investigations where the time-evolution contour is extended to the imaginary time axis. 
Only then it is possible to determine the temperature dependence of the excitation spectra which might serve as a thermometer for bosons in optical lattices \cite{hoffmann-paper}.\\
We acknowledge financial support from the German Academic Exchange Service (DAAD) and from the German Research Foundation (DFG) within the Collaborative Research Center SFB/TR12 
\textit{Symmetry and Universality in Mesoscopic Systems}.
\end{document}